\pdfoutput=1
\documentclass[11pt]{article}

\usepackage[preprint]{acl}

\usepackage{times}
\usepackage{latexsym}

\usepackage[T1]{fontenc}
\usepackage[utf8]{inputenc}

\usepackage{microtype}

\usepackage{inconsolata}

\usepackage{graphicx}

\usepackage{multirow}

\title{Wikimedia data for AI: 
a review of Wikimedia datasets for NLP tasks and AI-assisted editing}

\author{Isaac Johnson \\
  Wikimedia Foundation \\
  United States \\
  \texttt{isaac@wikimedia.org} \\\And
  Lucie-Aim\'{e}e Kaffee \\
  Hugging Face \\
  Germany \\
  \texttt{lucie.kaffee@huggingface.co} \\\And
  Miriam Redi \\
  Wikimedia Foundation \\
  United Kingdom \\
  \texttt{mredi@wikimedia.org}
  }

\begin{document}
\maketitle
\begin{abstract}
Wikimedia content is used extensively by the AI community and within the language modeling community in particular. In this paper, we provide a review of the different ways in which Wikimedia data is curated to use in NLP tasks across pre-training, post-training, and model evaluations. We point to opportunities for greater use of Wikimedia content but also identify ways in which the language modeling community could better center the needs of Wikimedia editors. In particular, we call for incorporating additional sources of Wikimedia data, a greater focus on benchmarks for LLMs that encode Wikimedia principles, and greater multilingualism in Wikimedia-derived datasets.
\end{abstract}

\section{Introduction}
Wikimedia data---especially Wikipedia---has been essential to the progression of AI over the past several years. In particular, Wikipedia text is key to natural language processing (NLP): it is generally long-form (meaning lots of context to learn from), ``well-written'',\footnote{\url{https://en.wikipedia.org/wiki/Wikipedia:Manual_of_Style}} and high-quality \cite{gao2020pile}. The BERT language model~\cite{devlin2018bert} that was introduced in 2018 and is often considered the first modern LLM uses English Wikipedia as a majority of its data. Even today, with much larger language models, English Wikipedia is often weighted heavily when trained---e.g., \cite{brown2020language,longpre2024pretrainer}.

The usage of Wikimedia data for AI has both been beneficial as a source of high-quality data for NLP researchers and for directing attention to the Wikimedia projects. This relationship, however, has largely been incidental to Wikimedia's mission and openness, and many of the advances of NLP have not made it back to the Wikimedia projects. For example, the Wikimedia Foundation regularly publishes snapshots of the content on the Wikimedia projects.  These ``dumps'' have been made available since at least 2005.\footnote{\url{https://meta.wikimedia.org/w/index.php?title=Data_dumps&oldid=216530}} While researchers have long been considered an expected end-user, this data was not pre-processed in any way to support the NLP community. As a result, researchers used many different approaches for pre-processing this raw text to produce natural-language text for use in training models.\footnote{See \citet{johnson2022considerations} for examples.} More recently, there have been explicit efforts to bring the Wikimedia and the ML communities closer together such as the Wiki-M3L\footnote{\url{https://meta.wikimedia.org/wiki/Wiki-M3L}} and NLP for Wikipedia\footnote{\url{https://meta.wikimedia.org/wiki/NLP_for_Wikipedia_(EMNLP_2024)}} workshops, and standardized datasets such as Hugging Face's Wikipedia text,\footnote{\url{https://huggingface.co/datasets/wikimedia/wikipedia}} There have also been concerns that the AI ecosystem might be depleting the very projects upon which it is built and stronger calls for developers of AI tools to view the knowledge commons not just a repository from which to extract data, but as a community to give back to---e.g., \citet{creativecommonsMakingWork} and \citet{wikimediaArtificialIntelligenceBellagio}.

In this paper, we make an effort to catalog the many AI and NLP-related datasets that draw on the Wikimedia projects to identify what gaps and opportunities exist. We frame this review following the calls for AI developers to contribute more to the knowledge commons. Specifically, we select the datasets in this paper with a focus on how NLP might be made more beneficial for the Wikimedia editor communities. Editors not only do the difficult work of synthesizing sources into the encyclopedic text consumed by readers and AI alike, they also engage in rich discussion and sense-making around source reliability, fairly portraying content, and evaluating complex questions of notability. Their work is guided by core content policies, which AI models must also be able to adhere to in order to be useful to the editing community. In the course of the analysis, we identify three major opportunities:
\begin{itemize}
    \item Extend and diversify the subset of Wikimedia data used in AI research. This could include regular datasets of images along with associated captions for multimodal modeling, more attention paid to talk pages or other collaboration spaces on Wikipedia, and greater usage of the high-quality transcribed documents produced by Wikisource communities.
    \item Consider the needs of Wikimedia editors in evaluation of LLMs. While Wikipedia data is well-represented within common benchmark datasets, these tasks are almost exclusively oriented towards reader goals. Work is needed to extend benchmarks to better encode the needs of Wikimedia editors.
    \item Continue to extend models to be more multilingual, open-source, and compact to meet the needs of the Wikimedia projects.
\end{itemize}

\section{Approach}
To guide the knowledge of Wikimedia datasets and tasks that are relevant to this work, we searched for individual Wikimedia projects on Hugging Face's dataset search\footnote{\url{https://huggingface.co/datasets}} and relied heavily on the authors' long experience working with Wikimedia data, developing natural language technologies, and collaborating with the Wikimedia communities. We list characteristic (not all) datasets for each stage of training, focusing on datasets and tasks that are oriented back towards the Wikimedia projects and that are at most a decade old. While we made an effort to build an exhaustive list, given the highly distributed nature of the Wikimedia movement and its research community, this overview might present some gaps. However, we believe that such potential gaps should not  affect our conclusions in major ways, and we treat this as the start of a catalog that we will work to update as we learn more and more datasets are created.

The current training paradigms of LLMs depend on datasets at three major stages: pre-training, post-training, and evaluation.\footnote{See \cite{dubey2024llama} for a good rundown of pre-training/post-training and \citet{bowman2021will} for a good overview of evaluation of LLMs.} Across these three stages, we detail how raw data is converted into datasets, tasks, and benchmarks to support the objectives of each stage. We see data, like the Wikimedia dumps, as relatively raw versions of what appears on the Wikimedia projects but in a form that is not directly useful for language models. We define datasets as data that has undergone pre-processing to anticipate a specific need, such as cleaning text to bring it closer to natural language. This pre-processing is important for turning Wikimedia content into high-quality datasets for language models to learn basic patterns of language (pre-training). We define tasks as datasets with explicit inputs and outputs that can be used to fine-tune models to complete a given action (post-training). In the final stage, benchmarks are curated tasks that allow for easy comparison of models to determine their usefulness to the Wikimedia projects (evaluation). While Wikimedia data has long been available and researchers have developed many datasets and tasks from this data, Wikimedia benchmarks have received less attention but are also an important mechanism for enabling members of the Wikimedia NLP community to encode our expectations for language models that are using Wikimedia content.

\subsection{What makes a dataset helpful to Wikimedia?}
\label{wmprinciples}
There are many many datasets that use Wikimedia data but not all of them relate to tasks that are clearly of value to the Wikimedia editor community.\footnote{We focus here on editors, but there are many other contributors to the Wikimedia projects that are also valuable stakeholders for future consideration such as campaign organizers or tool developers.} For example, SQuAD~\cite{rajpurkar2016squad} is a Q\&A dataset that is derived from Wikipedia that has played important role within the NLP community, but Q\&A does not necessarily map to a task where AI could directly help Wikimedia editors. While different editors and communities will have different needs, we highlight a few core principles that guide these needs and would ideally be expressed in datasets and the resulting models trained on Wikimedia data:
\begin{itemize}
    \item \textbf{Multilinguality}: Wikipedia alone exists in over 300 languages and providing equitable support to these different communities means building NLP tools that can handle their diversity.
    \item \textbf{Core content policies}: editors follow three core content policies\footnote{\url{https://en.wikipedia.org/wiki/Wikipedia:Core_content_policies}} that guide content on Wikipedia and useful models would need to do the same: Neutral Point-of-View (NPOV; fair representation of significant viewpoints), Verifiability (citations), and No Original Research (do not reach conclusions beyond the reliable sources).
    \item \textbf{Openness}: ``free'' and ``open'' are important to Wikimedia in many ways.\footnote{\url{https://w.wiki/B5zh}} In this context, language models are most useful when they are open-source and small enough to be reasonably hosted by the community, e.g., through the non-profit Wikimedia Foundation.
\end{itemize}

\subsection{From data to benchmarks: a case study}
Wikipedia articles offer an illustrative example of how data can be curated to support the three stages of training while adhering to the principles listed above. Starting with raw data, regular snapshots of the content of the Wikimedia projects have long been available as freely-downloadable dumps of article wikitext (the markup language used to write Wikipedia articles). For these dumps to be useful for most natural language applications---i.e. converted from raw data into a dataset---researchers both need to apply some basic filtering at the page level to remove non-content pages such as redirects and strip out the wikitext syntax from the pages to leave something closer to natural language.\footnote{See \citet{guo2020wiki} for an illustrative example.} These resulting natural language datasets are useful for pre-training but still require the identification of specific inputs and outputs to be converted into a task that can be used for post-training. As an example of post-training, \citet{qian2023webbrain} explore the task of writing short articles using an extensive dataset of Wikipedia article titles as inputs and the cleaned article text as expected outputs. Their metrics for automatic evaluation of the generated articles focus on language fluency and factualness. While this work is valuable for NLP fields like knowledge-intensive Q\&A, it only briefly explores metrics that capture Wikimedia principles such as Verifiability (appropriate citations). This makes the work less useful to the Wikimedia community as a benchmark that could allow for direct comparison of LLMs at assisting Wikimedians in producing high-quality content.

In contrast, FreshWiki~\cite{shao2024assisting} more directly aims to be this benchmark: it is a curated dataset of English Wikipedia articles that have been assessed to be of high quality (more likely to adhere to Wikimedia content policies) and that have been written largely after a specific cut-off date (to avoid data leakage due to memorization of Wikipedia content by LLMs). FreshWiki further incorporates citations in the expected output and adds metrics to measure how faithful the content is to its citations (see Table~\ref{tab:benchmarks}). While FreshWiki currently only exists in English, this same process could be extended to other language editions as there is nothing English-specific about it. \citet{shao2024assisting} evaluate GPT-4's performance, which is neither compact nor open-source, on FreshWiki because \cite{gao2023enabling} had previously shown that more open models (LLaMA-2 70B) performed well at generating text but still lagged behind GPT-4 in terms of correctly citing sources. Altogether, FreshWiki was able to better model Wikipedia's core content policies but exposed gaps in open models in this domain and is a framework that can be easily extended to be more multilingual.

\section{A review of curated Wikimedia data}
\subsection{Pre-training: from data to datasets}
\label{sec:pretraining}
Pre-training datasets for language models are collections of unsupervised text---i.e. no explicit task associated with them – that can be used to train language models to understand the basic relationships between words (tokens).\footnote{While we focus on language models, we also include some image-text data here.} These datasets are maximally useful when they are large, high-quality, and diverse. Datasets of Wikipedia articles are the prime example of this but they are not the only source of pre-training datasets available from the Wikimedia projects. Here, the needs of the Wikimedia projects are generally well-aligned with the needs of NLP researchers: better pre-training data means better models which can then be used to support the Wikimedia projects.

\begin{table*}[t]
  \centering
  \begin{tabular}{|p{0.3\linewidth}|l|p{0.4\linewidth}|}
    \hline
    Major source of text & Data available? & Pre-processed dataset? \\
    \hline
    Wikipedia articles & Various dumps\footnote{All dumps can be found at \url{https://dumps.wikimedia.org/}} & Hugging Face\footnote{\url{https://huggingface.co/datasets/wikimedia/wikipedia}} \\
    \hline
    Wikimedia Talk pages  & Various dumps & One-offs such as WikiConv~\cite{hua2018wikiconv} \\
    \hline
    Commons Images + captions / alt-text    & None &  One-offs such as WIT~\cite{srinivasan2021wit} or Concadia~\cite{kreiss2022concadia} \\
    \hline
    Wikisource transcriptions  & Various dumps & Hugging Face\footnote{\url{https://huggingface.co/datasets/wikimedia/wikisource}}      \\
    \hline
    Wikisource image-transcription pairs & None & None  \\
    \hline
    Other Wikimedia projects (Wikibooks, Wikivoyage, Wikiversity, Wiktionary) & Various dumps & None \\
    \hline
  \end{tabular}
  \caption{Major data(sets) of Wikimedia content.}
  \label{tab:pretraining}
\end{table*}

We distinguish here between whether data (raw content) is available and if there are standard datasets (pre-processed text). Table~\ref{tab:pretraining} shows two clear gaps: 1) raw data about image pixels and their associated text for pre-training of multimodal models is lacking, and, 2) even when the raw data is available, it is rare that standardized, pre-processed datasets are available that lower the barrier to access for researchers.\footnote{While this reduced barrier to entry feels appropriate for pre-training given that Wikipedia content is freely-licensed, we do encourage researchers to understand more deeply the content and processing choices that they are making when it comes to post-training.}

We encourage continued work to identify good practices for converting the other data sources listed in Table~\ref{tab:pretraining} into datasets. Each content source will bring its own challenges but the popularity of the Hugging Face Wikipedia dataset proves its value.\footnote{Over 100,000 downloads in August 2024 per \url{https://huggingface.co/datasets/wikimedia/wikipedia}.} For example, Wikisource offers an exciting opportunity to diversify the knowledge on which language models are being trained given the contributions by the Wikisource communities in digitizing knowledge from languages that have historically been underrepresented online.\footnote{\url{https://w.wiki/4Q7z}} Generating image datasets\footnote{Or e.g., audio transcriptions \cite{gomez2023speech}} will take much more work and resources given the massive size of the imagery hosted on Wikimedia Commons but would be a worthy addition to the outsized role that Wikimedia content plays in pre-training datasets.

One very positive aspect of the state of Wikimedia content for pre-training is that all of the data and almost all of the datasets are massively multilingual. While each of Wikipedia's over 300 language editions has varying norms and content, tools for converting this data into datasets generally are language-agnostic---i.e. they are stripping out syntax or making other choices that do not rely on tokenization or language-specific semantics. This helps to fuel a positive feedback loop of more multilingual content leading to more multilingual AI and thus more support for growing these language editions \cite{costa2022no}. As will be seen below, this wealth of language data unfortunately does not always hold for post-training datasets.

\subsection{Post-training: from datasets to tasks}
\label{sec:posttraining}
Post-training datasets for language models are collections of supervised tasks that can be used to fine-tune models to be more useful for end-users. Traditional fine-tuning converts a model from general language modeling to accomplishing a specific task that leverages a model's pre-trained language capabilities. Most LLMs are now instruction-tuned to not do any specific task but be generally capable of accomplishing many types of tasks.\footnote{Though traditional fine-tuning and instruction tuning have important differences in construction, we do not distinguish between the two as we generally believe that the datasets can be converted between the two formats as necessary.}

Below, we catalog these fine-tuning tasks with the goal of showing how Wikimedia content can be valuable in post-training and encouraging development of models that are more useful for Wikimedia-relevant tasks. Arguably the most salient usage of Wikimedia content for language modeling is related to Q\&A tasks---e.g., SQuAD \cite{rajpurkar2016squad} or WikiQA \cite{yang2015wikiqa}. Q\&A is a reader-focused task and one that receives plenty of attention in language modeling. Here we choose to focus on the needs of Wikimedia editors. In this domain, we see ample opportunity for LLM developers to make greater use of these Wikimedia-based post-training tasks. This would be beneficial for Wikimedians but should also support the general alignment goals of LLM developers as we will discuss in Section~\ref{sec:benchmarks}.

There are many possible transformations of Wikimedia data into post-training tasks. We represent this diversity by selecting a sample of tasks and example datasets for each one. We further split the tasks into three categories (classification, recommendation, and text generation) to provide some basic structure.

\paragraph{Classification}
\begin{itemize}
    \item \textbf{Stance detection}: a core part of Wikimedia is reaching consensus through discussions. \citet{kaffee2023should} studied article deletion discussions in English, German, and Turkish and fine-tuned a language model to predict what policies an editor will cite and their stance regarding deletion based on their comments.
    \item \textbf{Vandalism detection}: patrolling recent edits for vandalism that should be removed is a core task in maintaining Wikipedia's reliability. \citet{trokhymovych2023fair} fine-tuned language models in 47 languages to predict whether an edit will be reverted.
    \item \textbf{Citation-needed}: the Verifiability policy requires that many statements on Wikipedia be supported with a citation to a reliable source. \citet{redi2019citation} trained language models to predict whether a given sentence needs a citation in English, French, and Italian.
    \item \textbf{Readability}: accessibility of content to readers is important on Wikipedia but can be difficult to measure. \citet{trokhymovych2024open} fine-tuned language models in 14 languages to rank content by its readability.
    \item \textbf{NPOV detection}: a core content policy for Wikipedia is that text must adhere to a neutral point of view. \citet{wong2021wiki} built a dataset from English Wikipedia of edits that violated various policies for training classifiers to detect NPOV violations and other related content reliability issues.
\end{itemize}

\paragraph{Recommendation}
\begin{itemize}
    \item \textbf{Citation recommendation}: finding a source to verify a claim on Wikipedia can be a difficult task for editors. \citet{petroni2023improving} trained a retrieval and ranking model to find citations for statements on English Wikipedia.
    \item \textbf{Entity linking}: a key part of Wikipedia is its network of links that connect content and allow readers to go down rabbit holes. \citet{gerlach2021multilingual} trained a model across six language editions of Wikipedia for recommending links to be added to text spans within articles. There are also multimodal variants of this task such as visual entity linking.\footnote{\url{https://huggingface.co/datasets/aiintelligentsystems/vel_commons_wikidata}}
    \item \textbf{Grammatical error correction}: Fixing small spelling mistakes or grammatical errors is a common editing task on Wikipedia. \citet{grundkiewicz2014wiked} used English Wikipedia revision histories to identify these copy-edits in order to train language models for grammatical error correction.
\end{itemize}

\paragraph{Text Generation}
\begin{itemize}
    \item \textbf{Article descriptions}: all articles can be associated with a short phrase that helps readers disambiguate between similarly-named pages. \citet{sakota2023descartes} fine-tuned a language model to generate these article descriptions based on the first paragraph of Wikipedia articles and descriptions in other languages for 25 different language editions.
    \item \textbf{Edit Summaries}: each edit on Wikipedia should be accompanied by a short summary that explains what the edit did and why (similar to a code commit message). \citet{vsakota2024edisum} fine-tuned a language model to generate these edit summaries based on extracted diffs of a given edit on English Wikipedia.
    \item \textbf{Between Structured and Unstructured}: Facts can be stored in many different ways on the Wikimedia projects ranging from unstructured text in Wikipedia articles to semi-structured text in infoboxes or tables to the structured statements of Wikidata. Likewise, external sources of content to be incorporated can also be found in a variety of formats. Models for converting between these formats help editors in adding content and making it more accessible. For example, \citet{chen2021wikitablet} trained language models to produce long-form text from tabular data compiled from English Wikipedia while \citet{luggen2021wiki2prop} trained language models to recommend Wikidata properties based on Wikipedia text.
    \item \textbf{Natural language to SPARQL}: Wikidata contains a wealth of information but querying that content via what's known as SPARQL can be difficult. \citet{liu2024spinach} compile a dataset of English-language requests for SPARQL queries and the resulting query to evaluate LLM-based approaches for generating SPARQL queries.
    \item \textbf{Simplification}: Entire language editions (Simple English) and namespaces (Txikipedia) have been created on Wikipedia to provide simpler-language versions of content. \citet{sun2021document} use this correspondence between English and Simple English Wikipedia to build a dataset of article leads and their simpler equivalents to train language models to simplify text.
    \item \textbf{Summarization}: summarization has many potential use-cases on the wikis from helping editors understand long discussions on-wiki such as RFCs~\cite{im2018deliberation} or the information across multiple external sources. \citet{ghalandari2020large} compile a dataset from the English Wikipedia Current Events portal of multi-document summaries.
    \item \textbf{Machine translation}: translation plays an increasing role in assisting in content creation on Wikipedia and making the 300+ language editions accessible to all readers.\footnote{\url{https://www.mediawiki.org/wiki/MinT}} There are both datasets of published translations\footnote{\url{https://www.mediawiki.org/wiki/Content_translation/Published_translations}} for all languages and datasets of aligned text across languages like \citet{schwenk2021wikimatrix}.
    \item \textbf{Article writing}: Wikipedia is a tertiary source whose content is a consolidation of other sources as reflected in the citations. \citet{shao2024assisting} prompted LLMs to write English Wikipedia articles by gathering and summarizing sources related to a given topic.
\end{itemize}

This catalog of tasks demonstrates the diversity of NLP post-training tasks that already exist that could be beneficial to Wikimedia editors---ranging from simple binary classification to natural language generation, from short-form texts to long-form articles, and from models that must reflect Wikimedia-specific policies to more generic tasks like translation or summarization. This catalog also reveals large language gaps: despite the over 300 language editions of Wikipedia, most example datasets leverage English Wikipedia alone. This sometimes seems to be purely about precedent and familiarity---e.g., edit summaries exist in all language editions so expanding a dataset of them is largely trivial, but many language modeling tasks start with English. Other times, this stems from structural challenges on the Wikimedia projects that would take more extensive work to overcome---e.g., many language editions use various content reliability templates to flag NPOV issues but the templates and norms around them can vary language-to-language, making it difficult to scale datasets to more languages \cite{johnson2022considerations}.

We focused here on language as the most salient facet of these datasets, but as identified in Section~\ref{wmprinciples}, open-source licensing and compactness are also important to assessing the value of models to the Wikimedia projects. This is especially true in models that touch on privacy-sensitive areas such as search queries (e.g., natural language to SPARQL) where depending on 3rd-party models would open up individuals to surveillance. The NLP community has made important strides in both of these spaces in recent years but cataloging which tasks are lacking in good open-source models would be beneficial for considering future research.

\subsection{Evaluation: from tasks to benchmarks}
\label{sec:benchmarks}
Paraphrasing \citet{bowman2021will}, benchmarks for natural-language understanding are datasets that have the following characteristics: 1) they are representative of the task in question, 2) their data are accurate and unambiguous, 3) they can accurately rank models, and, 4) they disincentivize biased or harmful models. While the existence of many Wikimedia-focused tasks in Section~\ref{sec:posttraining} is heartening, few of these meet the standards of benchmarks. Trivially, many datasets that are derived from Wikimedia data can be found in the pre-training data used by many LLMs and thus are not accurate evaluations of these model's ability to generalize to new examples. This lack of Wikimedia benchmarks means that editors do not have easy or effective means of evaluating models (especially LLMs) for their usefulness to Wikimedia. Additionally, many LLMs are not open-source or are too large to be trained (or even fine-tuned in some cases) by Wikimedia developers. Developing core Wikimedia benchmarks could provide an important means of nudging NLP practitioners to develop models that are more beneficial out-of-the-box for the Wikimedia projects.

\begin{table*}[t]
  \centering
  \begin{tabular}{|p{0.15\linewidth}|p{0.35\linewidth}|p{0.4\linewidth}|}
    \hline
    Content Policy & Context & Benchmark \\
    \hline
    \multirow{2}{\linewidth}{Verifiability} & Creating content: given a topic to generate content, does the model appropriate cite its sources? & FreshWiki for English, which uses the citation quality metrics from ALCE~\cite{gao2023enabling} \\
     \cline{2-3}
    & Evaluating content: given a statement, does it require a citation? & Citation Needed~\cite{redi2019citation} for English, French, and Italian \\
    \hline
    \multirow{2}{\linewidth}{No Original Research} & Creating content: given a topic to generate content, does the model hallucinate any claims? & WildHallucinations~\cite{zhao2024wildhallucinations} which covers English Wikipedia and English non-Wikipedia topics. \\
    \cline{2-3}
    & Evaluating content: given a claim and source, is the claim supported? & FEVER~\cite{thorne2018fever} for English      \\
    \hline
    \multirow{2}{\linewidth}{Neutral Point-of-View (biased language)} & Creating content: given a topic or sentence, can the model remove biased language? & \multirow{2}{\linewidth}{\cite{pryzant2020automatically} and then \cite{ashkinaze2024seeing} for a more recent evaluation of LLMs and English Wikipedia.}  \\
    \cline{2-2}
    & Evaluating content: given a sentence, can the model identify if it uses biased language? & \\
    \hline
    \multirow{2}{\linewidth}{Neutral Point-of-View (due weight)} & Creating content: given a topic, can a model fairly represent all reliable sources? & WikiContradict~\cite{hou2024wikicontradict} is the closest analog, which evaluates how well models handle the summarization of contradictory information. \\
    \cline{2-3}
    & Evaluating content: given an article, can a model determine if the content is fairly represented? & None \\
    \hline
  \end{tabular}
  \caption{Benchmark tasks for Wikipedia's core content policies.}
  \label{tab:benchmarks}
\end{table*}

When it comes to evaluation of language models, it is less clear that the needs of the Wikimedia projects and NLP practitioners are currently well-aligned. Instruction-tuned LLMs are generally designed for a few purposes as demonstrated by the benchmarks that the model developers choose to test their models on. For example, the Llama 3 models \cite{dubey2024llama} are described as being benchmarked in eight top-level categories: (1) commonsense reasoning; (2) knowledge; (3) reading comprehension; (4) math, reasoning, and problem solving; (5) long context; (6) code; (7) adversarial evaluations; and (8) aggregate evaluations. Most of these categories are relevant for chat-bots to better answer questions but only incidentally tell us how these models might handle tasks related to applying Wikimedia content policies when editing or performing content moderation tasks. 

The core content policies of Wikipedia that guide many of the post-training tasks in Section~\ref{sec:posttraining} have clear corollaries with the intentions of LLMs developers. Neutral Point-of-View aligns well with training models that are not biased or harmful.\footnote{\citet{longpre2024pretrainer} showed that including Wikipedia in pre-training data greatly decreases model toxicity.} No Original Research aligns well with the goal of reducing hallucinations. Verifiability is perhaps less clear as a stated goal of many LLM models---i.e. the ability to cite sources for answers. However, we are witnessing a shift towards attribution of sources in LLM-backed products via retrieval-augmented generation, and Verifiability has nice overlap with chain-of-thought approaches \cite{khalifa2024source} that have been demonstrated to improve model performance in many reasoning tasks \cite{wei2022chain}. In all, LLMs that are more useful for Wikimedia-related tasks should also be more useful for many tasks outside of Wikimedia. In Table~\ref{tab:benchmarks}, we focus on these core content policies and examine the state of benchmarks for following these policies when creating content\footnote{Editing existing content is a different task but we also consider it under content creation.} as well as evaluating existing content for whether it adheres to the policy.

Table~\ref{tab:benchmarks} shows that there are existing benchmarks for evaluating the Verifiability and No Original Research policies. While citation-needed was developed with Wikipedia in mind, ALCE, FEVER, and WildHallucinations\footnote{WildHallucinations also covers content outside of Wikipedia but a related benchmark FActScore~\cite{min2023factscore} is extracted purely from English Wikipedia.} were developed with Wikipedia content but are oriented towards standard NLP tasks such as Q\&A or textual entailment. Work is still required to raise the quality of these benchmarks to ensure their freshness akin to FreshWiki's approach of only extracting content that was extensively edited after a given knowledge cut-off. And as with post-training tasks, these benchmarks are still heavily English-focused and do not cover the many other languages of Wikipedia.

Neutral Point-of-View has more mixed coverage. The NPOV policy contains multiple facets, of which two core components are the issue of biased language and the issue of biased coverage (due weight). Benchmarks do currently exist for the biased language facet based on editor activity from English Wikipedia. Biased coverage is harder to assess. WikiContradict\cite{hou2024wikicontradict} assesses a particular case where two reliable sources present contradictory information but there is a need for benchmarks that could e.g., determine whether content produced via multi-document summarization gives appropriate weight to different claims based on the level of their support across the documents. A core challenge here is not giving undue weight to fringe theories that may be mentioned by sources but are not well-supported.

We focused in this paper on the core content policies as an important first step for capturing facets important to the Wikimedia community and the basic existence of reasonable benchmarks in these areas. Moving forward, this framework could be extended to include more Wikimedia policies and guidelines and explore the fourth criteria asserted by \citet{bowman2021will} of disincentivizing bias through these benchmarks.

We recommend a few additional policies to consider for extending this framework.\footnote{We have linked to English Wikipedia policies and guidelines here but other language editions have developed their own policies and guidelines \cite{hwang2022rules}.} The policy on Copyright Violations\footnote{\url{https://en.wikipedia.org/wiki/Wikipedia:Copyright_violations}} touches on the importance of summarizing sources instead of copying them. Notability\footnote{\url{https://en.wikipedia.org/wiki/Wikipedia:Notability}} is a major guideline for determining whether an article should exist or not for a topic. Benchmarks might focus on evaluating sources for whether there is significant coverage of a given topic. There are also many style-related guidelines such as the Manual of Style\footnote{\url{https://en.wikipedia.org/wiki/Wikipedia:Manual_of_Style}} which touch on how to structure and format content such as capitalization, abbreviations, and mixing of dialects. One gap that is unlikely to be filled is assessing source reliability (a core component of all three core content policies). English Wikipedia, for example, tracks sources whose reliability is often questioned in a list known as Perennial Sources\footnote{\url{https://en.wikipedia.org/wiki/Wikipedia:Reliable_sources/Perennial_sources}}. These assessments can change as sources themselves evolve and reflect consensus from long discussions about these sources. It is both hard to imagine LLMs making these assessments (except perhaps as a support for summarizing discussions) and undesirable to leave this complex sense-making to AI.

For disincentivizing bias through benchmarks, there is a long history of research on biases on the Wikimedia projects to pull from \cite{redi2020taxonomy}. One key step is expanding benchmarks to cover more languages but researchers might also develop benchmarks that only use articles that comprise a more balanced representation of the world. Datasets like \citet{merity2016pointer} that filter articles to only those deemed to be of the highest quality by Wikimedians would be another way to ensure that benchmark data is maximally likely to e.g., fully meet the expectations of the NPOV policy.

\section{Conclusion}
We present a summary of how Wikimedia data is curated to support the different stages of model training with a focus on NLP. At each stage, we highlight data that could be converted into more useful forms for training language models and identify ways in which these models could be more useful for Wikimedia editors. This shows that while Wikimedia content has been hugely influential and important to the development of AI as a source of language data, the field still has gaps in developing benchmarks and models that reflect the needs of Wikimedia editors. We hope that the opportunities that we highlight in this space encourage a more mutualistic relationship between NLP and the Wikimedia communities.

\bibliography{custom}

\begin{table*}[t]
  \centering
  \begin{tabular}{|p{0.45\linewidth}|p{0.25\linewidth}|p{0.2\linewidth}|}
    \hline
    Task & Dataset & Language Coverage \\
    \hline
    \multicolumn{3}{|c|}{Classification} \\
    \hline
    \textbf{Stance detection}: predicting positions in consensus discussions & \citet{kaffee2023should} & English, German, and Turkish  \\
    \hline
    \textbf{Vandalism detection}: predicting whether an edit should be reverted. &  \citet{trokhymovych2023fair} & 47 languages \\
    \hline
    \textbf{Citation-needed}: predicting whether a sentence needs a citation. & \citet{redi2019citation} & English, French, and Italian \\
    \hline
    \textbf{Readability}: measuring accessibility of content to readers  & \citet{trokhymovych2024open} & 14 languages    \\
    \hline
    \textbf{NPOV detection}: predicting whether a statement uses biased language. & \citet{wong2021wiki} & English \\
    \hline
    \multicolumn{3}{|c|}{Recommendation} \\
    \hline
    \textbf{Citation recommendation}: finding the best source to verify (or refute) a claim on Wikipedia. & \citet{petroni2023improving} & English \\
    \hline
    \textbf{Entity linking}: recommending links to be added to Wikipedia articles.\footnote{There are also multimodal variants of this task such as visual entity linking: \url{https://huggingface.co/datasets/aiintelligentsystems/vel_commons_wikidata}} & \citet{gerlach2021multilingual} & Six languages \\
    \hline
    \textbf{Grammatical error correction}: recommending fixes to small spelling mistakes or grammatical errors. & \citet{grundkiewicz2014wiked} & English \\
    \hline
    \multicolumn{3}{|c|}{Text Generation} \\
    \hline
    \textbf{Article descriptions}: generating a short phrase to summarize the topic of an article. & \citet{sakota2023descartes} & 25 languages \\
    \hline
    \textbf{Edit Summaries}: describing what an edit did on Wikipedia and why. & \citet{vsakota2024edisum} & English \\
    \hline
    \textbf{Structured to Unstructured}: converting from tables or Wikidata statements to natural language. & \citet{chen2021wikitablet} & English \\
    \hline
    \textbf{Unstructured to Structured}: converting from Wikipedia or textual sources to more structured tables or Wikidata statements. & \citet{luggen2021wiki2prop} & English, French, and German \\
    \hline
    \textbf{Natural language to SPARQL}: converting from a user question to a Wikidata query & \citet{liu2024spinach} & English \\
    \hline
    \textbf{Simplification}: reducing the complexity of text to make it more accessible to readers. & \citet{sun2021document} & English \\
    \hline
    \textbf{Summarization}: converting discussions or multiple sources into short summaries to judge their relevance. & \citet{im2018deliberation} (editor discussions); \citet{ghalandari2020large} (sources) & English \\
    \hline
    \textbf{Machine translation}: converting Wikipedia content from one language to another to assist editors or readers.\footnote{\url{https://www.mediawiki.org/wiki/MinT}} & \cite{schwenk2021wikimatrix}. & 96 languages \\
    \hline
    \textbf{Article writing}: writing appropriately-cited articles about a topic & \citet{shao2024assisting} & English \\
    \hline
  \end{tabular}
  \caption{Major datasets and associated tasks relevant to Wikimedia editors.}
  \label{tab:posttraining-simple}
\end{table*}

\end{document}